\documentclass[aps,prl,twocolumn,amsmath,amssymb, superscriptaddress]{revtex4-2}
\usepackage{graphicx}
\graphicspath{{Figures/}}
\usepackage{subfigure}
\usepackage{epsfig}
\usepackage{ulem}
\usepackage{dcolumn}
\usepackage{bm}
\usepackage{hyperref}
\hypersetup{colorlinks=true, citecolor=blue, urlcolor=blue, linkcolor=blue}
\def\beq{\begin{equation}}
\def\eeq{\end{equation}}
\def\bald{\begin{aligned}}
\def\eald{\end{aligned}}
\def\bea{\begin{eqnarray}}
\def\eea{\end{eqnarray}}

\def\ket#1{\left|#1\right\rangle}
\def\avg#1{\left\langle#1\right\rangle}

\def\Eq#1{Eq.~(\ref{#1})}
\def\Fig#1{Fig.~\ref{#1}}

\begin{document}
\title{Boosting quantum Monte Carlo and alleviating sign problem by Gutzwiller projection}
\author{Wei-Xuan Chang}
\affiliation{Beijing National Laboratory for Condensed Matter Physics and Institute of Physics,
Chinese Academy of Sciences, Beijing 100190, China}
\affiliation{University of Chinese Academy of Sciences, Beijing 100049, China}
\author{Zi-Xiang Li}
\email{zixiangli@iphy.ac.cn}
\affiliation{Beijing National Laboratory for Condensed Matter Physics and Institute of Physics,
Chinese Academy of Sciences, Beijing 100190, China}
\affiliation{University of Chinese Academy of Sciences, Beijing 100049, China}

\begin{abstract}
Here we develop a new scheme of projective quantum Monte-Carlo (QMC) simulation combining unbiased zero-temperature (projective) determinant QMC and variational Monte-Carlo based on Gutzwiller projection wave function, dubbed as ``Gutzwiller projection QMC''. The numerical results demonstrate that employment of Gutzwiller projection trial wave function with minimum energy strongly speed up the convergence of computational results, thus tremendously reducing computational time in the simulation. More remarkably, we present an example that sign problem is enormously alleviated in the Gutzwiller projection QMC, especially in the regime where sign problem is severe. Hence, we believe that Gutzwiller projection QMC paves a new route to improving the efficiency, and alleviating sign problem in QMC simulation on interacting fermionic systems.   
\end{abstract}
\date{\today}

\maketitle
{\bf Introduction:} Demystifying quantum many-body physics in strongly correlated systems is of central importance in modern condensed matter physics. Developing efficient numerical approaches to solve quantum many-body systems in more than one dimension is particularly crucial. Among various quantum many-body numerical algorithms, quantum Monte-Carlo (QMC) plays a vital role because it is unbiased and approximation-free\cite{Scalapino1981PRL,Hirsch1981PRL,White1989PRL,Zhang1995PRL,Gull2011RMP,Assaad2008world}. However, QMC suffers from the notorious sign problem\cite{Troyer2005PRL,ZXLiQMCreview,SignWhite,Mondaini2022Science}, which strongly hinders application of QMC simulation to many strongly correlated systems, for instance Hubbard model at generic filling. Hence, solving or alleviating sign problem in quantum many-body models potentially featuring intriguing physics will definitely lead to a substantial progress in understanding strongly correlated physics in quantum many-body systems\cite{Li2015PRB,Li2016PRL,Sorella2007PRL,Hangleiter2020ScienceAdvances,wan2020mitigating,Xu2022PRB,Xu2021PRB, Wang2015PRL,Xiang2016PRL,Clark2021PRL}

On the other side, for quantum Monte-Carlo simulation on interacting fermionic systems, the computational complexity is generally cubic in linear system size, which strongly hampers the applications to the fermionic systems with large system sizes\cite{Assaad2008world}. Despite the recent progress on the development of improved algorithms for fermionic QMC\cite{He2019PRL,Scalettar2022PRE,Wang2015PRB}, in general the QMC simulations on interacting fermionic systems are much more costly compared with the ones on spin or bosonic systems.  Hence, although in the simulation free from sign problem, it is extremely difficult to access the accurate properties close to the thermodynamic limit in studying some important issues, for instance quantum criticality\cite{Berg2019Review,Berg2012Science,Xu2019Review} and competing ordering\cite{Kivelson2015RMP}. Designing powerful algorithms for interacting fermionic models with high efficiency is 
thus immensely desirable.   

\begin{figure}[tb]
\includegraphics[width=0.52\textwidth]{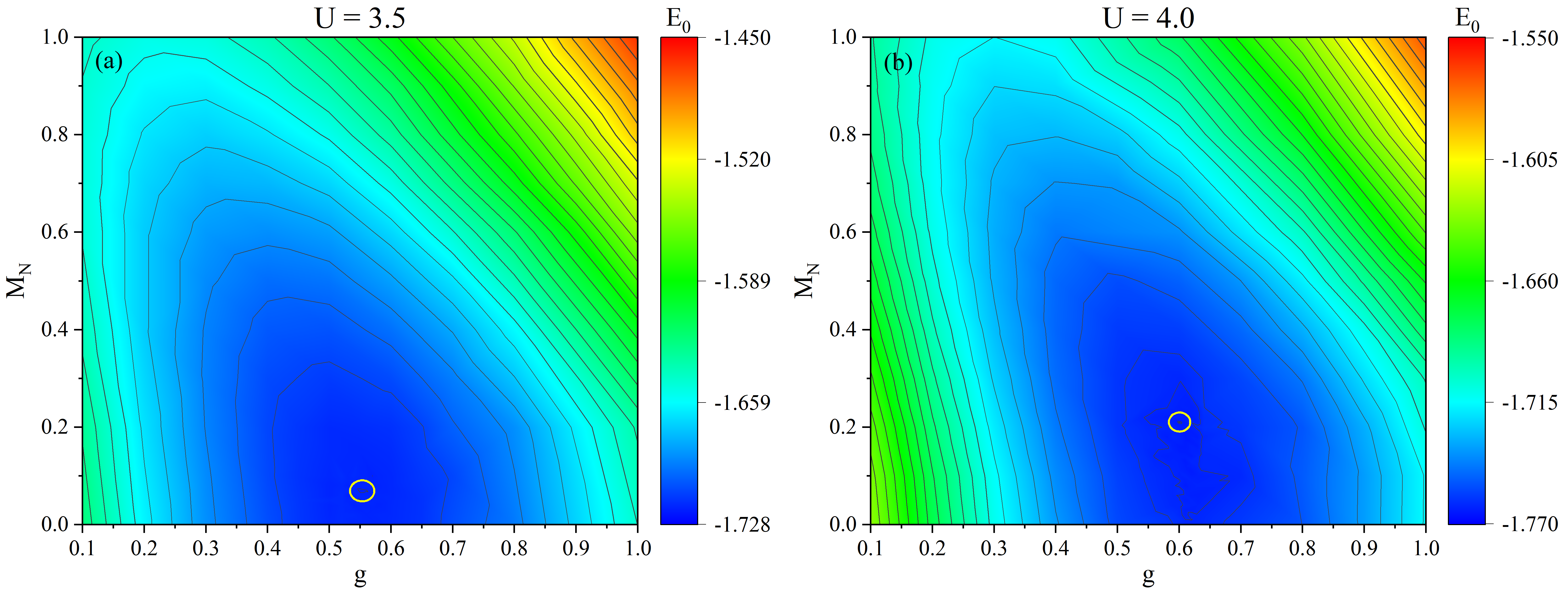}
\caption{ The contour plot of energy versus variational parameters in Gutzwiller projection wave function for $U=3.5$(a) and $U=4.0$(b). The parameters $(g,M_{\rm{N}})$ with minimum energy are indicated by the white circles. The optimal parameters are $g=0.55$ and $M_{\rm{N}}=0.07$ for $U=3.5$, and $g=0.6$ and $M_{\rm{N}}=0.21$ for $U=4.0$. }
\label{Fig1}
\end{figure}

To this end, we propose a new scheme of QMC simulation dubbed as ``Gutzwiller projection QMC'' to speed up the simulation, and more remarkably, alleviate sign problem in interacting fermionic models. The basic idea of the approach is the combination of variational Monte-Carlo based on the Gutzwiller projection variational wave function\cite{Gutzwiller1963PRL,Rice1986PRB, Kotliar1991PRB, Zhou2021PRB} and intrinsically unbiased projection QMC. We implement a mean-field wave function under Gutzwiller projection as the trial wave function, and employ standard procedure of projection QMC to access the ground-state of an interacting Hamiltonian without involving any uncontrolled approximations. In the framework of Hubbard-Strotonovich transformation utilized in the PQMC, the optimal Gutzwiller variational wave function with minimum energy is achieved efficiently. Compared with the conventional PQMC simulation which implements a slater-determinent trial wave function, much smaller projection parameter is required to guarantee the convergence of results in Gutzwiller projection QMC, hence immensely reducing the computational time. More crucially, the systematic calculations in specific models reveal that sign problem is greatly mitigated in the Gutzwiller projection QMC.

\begin{figure*}[t]
\includegraphics[width=1.0\textwidth]{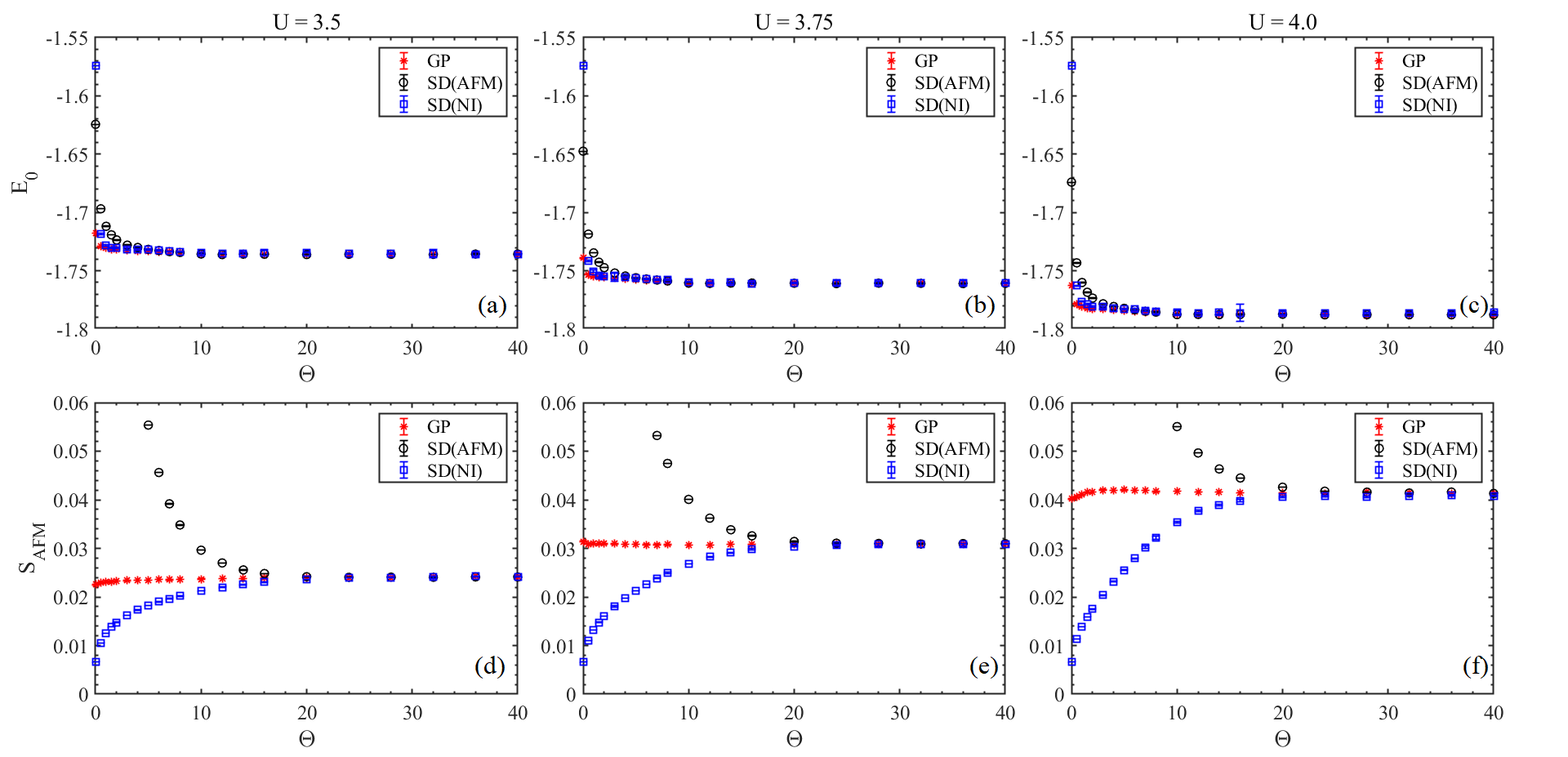}
\caption{The results of simulation on spinful Hubbard model at half filling for Gutzwiller QMC and conventional PQMC with slater-determinant trial wave function. The results of ground-state energy versus projective parameter $\Theta$ for (a) $U=3.5$, (b) $U=3.75$ and (c) $U=4$. The results of AFM structure factor $S_{AFM}$ versus projective parameter $\Theta$ for (d) $U=3.5$, (e) $U=3.75$ and (f) $U=4$. GP, SD(AFM), SD(NI) denote the results of simulation with Gutzwiller projective wave function, AFM mean-field slater-determinant wave function and non-interacting slater-determinant wave function as the trial wave functions, respectively.  }
\label{Fig2}
\end{figure*}

{\bf Method: } In this section, we briefly illustrate the methods of Gutzwiller QMC, combining variational Monte-Carlo based on Gutzwiller projection wave function and unbiased projection QMC. To avoid complexity, we consider the typical wave function with on-site Gutzwiller projection to illustrate our strategy: $\ket{\psi_G} = e^{-g \sum_i n_{i\uparrow}n_{i\downarrow}} \ket {\psi_{M}}$, where $\ket{\psi_{M}}$ is a slater-determant wave function featuring mean-field ordering, and $e^{-g n_{i\uparrow}n_{i\downarrow}}$ is Gutzwiller projection with projective parameter $g$. The first step is minimizing the expectation value of Hamiltonian under the given Gutzwiller projection variational wave function, which yields the optimal parameters of projective parameter $g$ and mean-field order parameter in slater-determinant wave function $\ket{\psi_{M}}$. Here we perform H-S transformation on the Gutzwiller projection term: 
\bea
e^{-g n_{i\uparrow}n_{i\downarrow}}=\frac{1}{2} e^{-g/4}\sum_{s_i =\pm 1}e^{\lambda s_i (n_{i\uparrow}+n_{i\downarrow})}
\label{Gutzwiller}
\eea
where $\cosh \lambda = e^{\frac{g}{2}}$ and $s_i = \pm 1$ is the auxiliary field defined on each site $i$. Then the expectation value $\langle\hat{O}\rangle=\frac{\langle\psi_G|\hat{O}|\psi_G\rangle}{\langle\psi_G|\psi_G\rangle}$ is straightforwardly achieved utilizing the standard procedure of determinant QMC. Notice that this approach of yielding the expectation value of observables based on Gutzwiller projection wave function is significantly faster than the conventional approach in variational Monte-Carlo. Then we search the optimal parameters in Gutzwiller projection wave function by minimizing the expectation value of energy.

Upon obtaining the optimal Gutzwiller projection wave function, we compute the expectation value of observable as $\langle\hat{O}\rangle=\frac{\langle\psi_T|e^{-\Theta \hat{H}}\hat{O} e^{-\Theta \hat{H}}|\psi_T\rangle}{\langle\psi^o_G|e^{-2\Theta \hat{H}}|\psi^o_G\rangle}$ with the trial wave function $|\psi_T\rangle = e^{-g \sum_i n_{i\uparrow}n_{i\downarrow}} \ket {\psi_{M}} $. We employ the standard procedure of Trotter decomposition and H-S transformation in PQMC, and decouple the Gutzwiller projection using \Eq{Gutzwiller}. Then the ground-state expectation values of observable are easily accessed in the scheme of conventional PQMC.

{\bf The honeycomb Hubbard model:} To demonstrate the efficiency of Gutzwiller QMC, we first apply the approach to spin-1/2 Hubbard model\cite{Kivelson2022Review} on honeycomb lattice:
\bea\label{Model1}
H=-t\sum_{\avg{ij},\sigma}(c^\dagger_{i\sigma}c_{j\sigma}+h.c.) + U\sum_{i} n_{i\uparrow}n_{i\downarrow},
\eea
where $c^\dag_{i\sigma}$ creates an electron on site $i$ with spin polarization $\sigma=\uparrow$/$\downarrow$, $t$ is the nearest-neighbor (NN) hopping, and $U$ is the amplitude of onsite Hubbard repulsion. Hereafter we set $t=1$ as unit of energy. We focus the study at half filling where sign problem is circumvent upon choosing appropriate HS transformation channel. In the non-interacting limit namely $U=0$, the model at half filling features Dirac fermions with Fermi energy located at Dirac point. With increasing Hubbard interaction amplitude, a quantum phase transition between Dirac semimetal and antiferromagnetic Mott insulator occurs $U\!=\!U_c\!\approx\! 3.85$ and the transition belongs to chiral-Heisenberg universality class\cite{sorella2016PRX,Herbut2014PRB,Janssen2023PRB,Herbut2017PRD,Assaad2015PRB}. 

We implement Gutzwiller QMC algorithm to study the ground-state properties of \Eq{Model1} at half filling. Because AFM is the dominant ordering in the model, it is natural to choose AFM mean-field wave function with Gutzwiller projection as the trial wave function in the simulation:
\bea
\ket{\psi_T} = e^{-g \sum_i n_{i\uparrow}n_{i\downarrow}} \ket {\psi_{N}}
\eea
where $g$ is the parameter of Gutzwiller projection, and $\psi_{\rm{N}}$ is mean-field wave function featuring Neel AFM ordering, more explicitly, the ground-state wave function of the Hamiltonian $H_{N} = H_{0} + M_{N} \sum_i (-1)^{\delta_i} (n_{i\uparrow} - n_{i\downarrow})$, where $H_{0}$ is the non-interacting part of \Eq{Model1}, $M_{\rm{N}}$ is Neel AFM order parameter, and $\delta_i =\pm 1$ if site $i$ belongs to A(B) sublattice. Employing the procedure introduced in the section of Method, it is straightforward to access the expectation value of \Eq{Model1} in terms of wave function under the choice of $g$ and $M_{\rm{N}}$. \Fig{Fig1} depicts the expectation value of energy with varying Neel order parameter $M_{\rm{N}}$ and Gutzwiller projection parameter $g$, for several choices of Hubbard interaction $U=3.5$ and $U=4.0$ located in the DSM phase and AFM ordered phase, respectively. We obtain the optimal values of parameters $g$ and $M_{\rm N}$ by minimizing the expectation value of \Eq{Model1}, and achieve the Gutzwiller variational wave function utilized in the Gutzwiller QMC simulation.

We compare the convergence of results against projection parameter $\Theta$ using distinct choices of trial wave function. The results of ground-state energy and AFM structure factor (The definition are shown in the Supplementary Materials) are evaluated at varying $\Theta$, as depicted in \Fig{Fig2}, which unambiguously show accurate ground-state energy is achieved at much smaller value of $\Theta$ in Guzwiller QMC, compared with conventional DQMC simulation which implements slater-determinant wave function as trial wave function. In the conventional QMC simulation, two different choices of slater-determinant trial wave function are employed, including the ground-state wave function of non-interacting part in \Eq{Model1} and AFM mean-field wave function in the absence of Gutzwiller projection. The results of ground-state energy and AFM structure factor $S_{\rm{AFM}}$ exhibit much slower convergence against projection parameter $\Theta$ in both cases. More surprisingly, within same number of Monte-Carlo sampling, the statistic errors of observable are significantly reduced in the Gutzwiller QMC, compared with the conventional DQMC with slater-determinant trial wave function. The corresponding results of statistic error are included in the Supplementary Materials.

\begin{figure}[tb]
\includegraphics[width=0.5\textwidth]{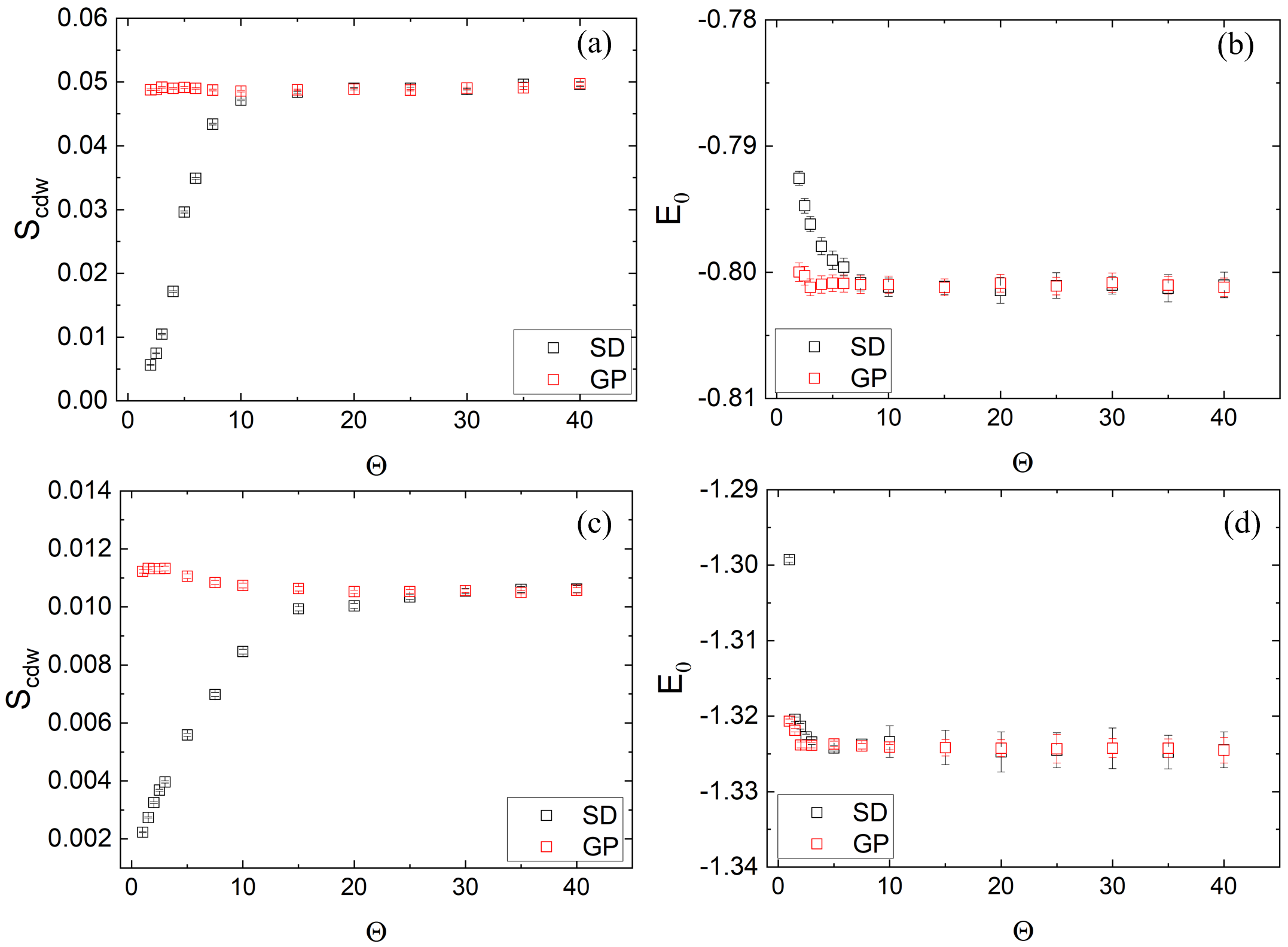}
\caption{The results of observable in spinless t-V model at half filling for Gutzwiller QMC and conventional PQMC with slater-determinant trial wave function. The results of CDW structure factors $S_{\rm{CDW}}$ for (a) V=1.6 and (c) V=1.35. The results of ground-state energy $E_0$ for (b) V=1.6 and (d) V=1.35. SD denotes the results of conventional PQMC with employment of non-interacting slater-determinant trial wave function, and GP denotes the results of Gutzwiller 
 QMC. }
\label{Fig3}
\end{figure}

\begin{figure*}[t]
\includegraphics[width=1.0\textwidth]{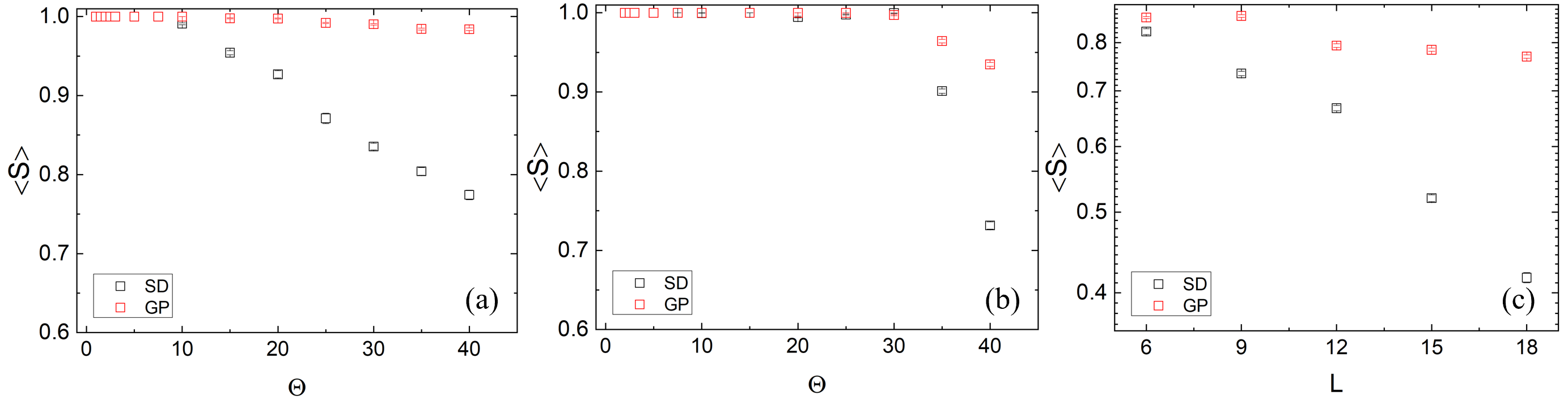}
\caption{The results of sign problem in spinless t-V model at half filling for Gutzwiller QMC and conventional PQMC with slater-determinant trial wave function. (a) The result of sign problem versus projective parameter $\Theta$ for $L=15$ and $V=1.6$. (b) The results of sign problem versus projective parameter $\Theta$ for $L=15$ and $V=1.35$. (c) The results of sign problem versus linear system size $L$ for $V=1.6$. SD denotes the results of conventional PQMC with the employment of non-interacting slater-determinant trial wave function, and GP denotes the results of Gutzwiller QMC.    }
\label{Fig4}
\end{figure*}

{\bf Repulsive spinless honeycomb model:} 
For Honeycomb Hubbard model, we unambiguously show that utilizing Gutzwiller trial wave function strongly expedite the convergence of results against projection parameter $\Theta$, resulting in a tremendous improvement of the efficiency in the QMC simulation. In this section, we apply Gutzwiller QMC to the spinless t-V model, namely spinless fermions model with nearest-neighbor (NN) interaction, on honeycomb lattice. The Hamiltonian of the model reads:
\bea\label{Model2}
H=-t\sum_{\avg{ij}}(c^\dagger_{i}c_{j}+h.c.) + V \sum_{\avg{ij}} (n_i-\frac{1}{2})(n_j-\frac{1}{2})
\eea
where $c_i$ is the annihilation operator of fermion on site $i$, $t$ is the NN hopping amplitude, and $V>0$ denotes the density repulsive interaction between NN sites. We focus on the half filling of the model. The quantum phase diagram of the model at half filling has been extensively investigated in recent years, which features a quantum phase transition from DSM to a charge-density wave (CDW) insulating phase with increasing NN density interaction strength\cite{Li2015NJP,Wang2014NJP}.  The appearance of sign problem depends on the schemes of HS transformation. The simulation is sign problematic if the NN density interaction is decoupled in the hopping channel. Recently, it is shown that although sign problem exists, the behaviour of average sign exhibit two distinct behaviours in the weak and strong coupling regimes\cite{Li2022arXiv}. In the weak coupling regime, the model is asymptotic sign-free, namely average sign asymptotically increases to one as system size increases, whereas in the strong coupling regime the average sign exhibit exponentially decaying scaling consistent with conventional scaling behaviour of sign problem in QMC simulation. Here, we decouple the interaction in the density channel and perform Gutzwiller QMC simulation on spinless t-V model at half filling, aiming at investigating the effect of Gutzwiller trial wave function on the efficiency of simulation, and more importantly, the behaviour of sign problem.

In the simulation of spinless honeycomb t-V model, we choose CDW mean-field wave function with Gutzwiller projection on the NN bond as the trial wave function: 
\bea
\ket{\psi_T} = e^{-g \sum_{\avg{ij}}n_{i}n_{j}} \ket {\psi_{C}}
\eea
where $g$ is the parameter of Gutzwiller projection on NN bonds. $\psi_{C}$ is a mean-field wave function with CDW ordering, generated as the ground-state wave function of the mean-field Hamiltonian $H_{\rm C} = H_0 + \Delta_{\rm C}\sum_i (-1)^{\delta_i} n_i$, where $H_0$ is the non-interacting part in Hamiltonian \Eq{Model2}, $\Delta_{\rm C}$ is CDW order parameter and $\delta_i =\pm 1$ if site $i$ belongs to A(B) sublattice. Similar to spinful Honeycomb Hubbard model, we access the optimal variational parameters $g$ and $\Delta_{\rm C}$ by minimizing expectation value of $H$ defined in \Eq{Model2}. We perform QMC simulation with the corresponding Gutzwiller trial wave function. The results of ground-state energy and CDW structure factors, with the detailed definitions included in Supplementary Materials, for several values of $V$ are presented in Fig.3, explicitly demonstrating that smaller value of $\Theta$ is sufficient to access the accurate ground-state results of energy and CDW structure factors in Gutzwiller QMC. The improvement of efficiency is particularly pronounced in the CDW ordered phase ($V>V_c\sim 1.35$).

Then we investigate the behaviour of sign problem in the Gutzwiller QMC simulation. For simulation of \Eq{Model2} with HS transformation in density channel, as aforementioned, the model is intrinsically sign-problematic only as interaction is strong $V>V^* \sim 1.2$, whereas the model displays asymptotic sign-free behaviour in the weak coupling regime, namely average sign increases and approaches to one as system size increases. Hence, we focus on the strong coupling regime where sign problem is severe. We calculate the average sign as a function of projection parameter $\Theta$ for $V=1.35$ and $V=1.6$, as depicted in \Fig{Fig4}(a) and \Fig{Fig4}(b), respectively. In comparison, we present the results employing conventional DQMC with slater-determinant trial function generated as the ground state of non-interacting part in \Eq{Model2}. Intriguingly, the results of average sign is obviously increased in the simulation with Gutzwiller trial wave function, compared with the ones in conventional DQMC with slater-determinant trial wave function, hence sign problem is significantly alleviated in Gutzwiller QMC. Furthermore, we plot the average sign versus linear system size in the simulation with Gutzwiller QMC and conventional QMC for $V=1.6$ (shown in \Fig{Fig4}(c)), which unequivocally demonstrates that sign problem in spinless honeycomb t-V model at half filling is tremendously mitigated with the employment of Gutzwiller trial wave function.

{\bf Discussions and concluding remarks:} In summary, we develop a new scheme of zero-temperature (projective) DQMC boosted by Gutzwiller projection. To demonstrate the remarkable efficiency, we apply the approach to two typical quantum many-body models. In spinful Honeycomb Hubbard model, the results clearly show that important observables including ground-state energy and AFM structure factors exhibit much faster convergence against projection parameter $\Theta$, thus resulting in tremendous reduction of needed computational time to achieve certain accuracy. In spinless Honeycomb t-V model, the convergence of observable is also achieved with smaller projection parameter in Gutzwiller QMC, compared with the conventional algorithm of projective DQMC. More appealingly, the sign problem of the model is significantly alleviated in Gutzwiller QMC simulation if the HS transformation is performed in sign-problematic channel, especially in the regime where sign problem is severe. To conclude, Gutzwiller QMC offers a potential route to speeding up the simulation on the ground-state properties on the interacting fermionic models, and alleviating sign problem in the simulation on the sign-problematic models.

\textit{Acknowledgement}: We would like to thank Yi Zhou and Hong Yao for helpful discussions. This work is supported in part by the
start-up grant of IOP-CAS.

\bibliography{Signproblem-bib}

\end{document}